\documentclass[12pt]{article}
\usepackage{amsfonts,amsmath,graphicx,setspace}
\usepackage{color,hyperref,verbatim,natbib,rotating}

\definecolor{Red}{rgb}{0.5,0,0}
\definecolor{Blue}{rgb}{0,0,0.5}
\hypersetup{%
    colorlinks = {true},
    linktocpage = {true},
    plainpages = {false},
    linkcolor = {Blue},
    citecolor = {Blue},
    urlcolor = {Red},
    pdfstartview = {XYZ null null 1.25},
    pdfpagemode = {UseOutlines},
    pdfview = {XYZ null null null}
}
\newcommand{\email}[1]{\href{mailto:#1}{\normalfont\texttt{#1}}}
\newcommand{\R}{\textsf{R}}

\newcommand{\bfalpha}{\mbox{{\boldmath $\alpha$}}}
\newcommand{\bfbeta}{\mbox{{\boldmath $\beta$}}}
\newcommand{\bfgamma}{\mbox{{\boldmath $\gamma$}}}
\newcommand{\bftheta}{\mbox{{\boldmath $\theta$}}}
\newcommand{\bflambda}{\mbox{{\boldmath $\lambda$}}}
\newcommand{\eps}{\varepsilon}
\newcommand{\bw}{\mbox{{\boldmath $w$}}}
\newcommand{\bv}{\mbox{{\boldmath $v$}}}
\newcommand{\by}{\mbox{{\boldmath $y$}}}
\newcommand{\bx}{\mbox{{\boldmath $x$}}}

\newcommand{\bz}{\mbox{{\boldmath $z$}}}

\newcommand{\bb}{\mbox{{\boldmath $b$}}}

\newcommand{\bD}{\mbox{{\boldmath $D$}}}
\newcommand{\aucHat}{A$\widehat{\mbox{U}}$C}
\newcommand{\ipeHat}{I$\widehat{\mbox{P}}$E}

\bibpunct{(}{)}{;}{a}{}{,}

\begin{document}

{\vspace*{1cm}}

\begin{center}
\Large \bf Dynamic Predictions with Time-Dependent Covariates in Survival Analysis using Joint Modeling and Landmarking
\end{center}

\vspace{0.5cm}
\begin{center}
{\large Dimitris Rizopoulos$^{1,*}$, Magdalena Murawska$^{1}$, Eleni-Rosalina Andrinopoulou$^{1,2}$, Geert Molenberghs$^{3}$, Johanna J.M. Takkenberg$^{2}$ and Emmanuel Lesaffre$^{1,3}$}\footnote{$^*$Correspondance to: Department of Biostatistics, Erasmus University Medical Center, PO Box 2040, 3000 CA Rotterdam, the Netherlands. E-mail address: \email{d.rizopoulos@erasmusmc.nl}.}\\\quad\\
$^{1}$ Department of Biostatistics, Erasmus Medical Center, the Netherlands\\
$^{2}$ Department of Cardiothoracic Surgery, Erasmus Medical Center, the Netherlands\\
$^{3}$ Interuniversity Institute for Biostatistics and statistical Bioinformatics,\\Katholieke Universiteit Leuven \& Universiteit Hasselt, Belgium\\
\end{center}
\vspace{0.6cm}


\begin{spacing}{1}
\noindent {\bf Abstract}\\
A key question in clinical practice is accurate prediction of patient prognosis. To this end, nowadays, physicians have at their disposal a variety of tests and biomarkers to aid them in optimizing medical care. These tests are often performed on a regular basis in order to closely follow the progression of the disease. In this setting it is of medical interest to optimally utilize the recorded information and provide medically-relevant summary measures, such as survival probabilities, that will aid in decision making. In this work we present and compare two statistical techniques that provide dynamically-updated estimates of survival probabilities, namely landmark analysis and joint models for longitudinal and time-to-event data. Special attention is given to the functional form linking the longitudinal and event time processes, and to measures of discrimination and calibration in the context of dynamic prediction. 
\end{spacing}


\section{Introduction} \label{Sec:Intro}
Nowadays there is great interest in accurate risk assessment for prevention and treatment of disease. Physicians use risk scores to reach appropriate decisions, such as prescribing treatment, or extra medical tests or suggesting alternative therapies. Patients who are informed about their health risk often decide to adjust their lifestyles to mitigate it. Risk scores are typically based on several factors that describe the patients' physical condition, such as age, BMI, smoking, genetic predisposition, and the results of medical tests. In this work we focus on the use of the results of such tests and more specifically on biomarkers. The majority of prognostic models in the medical literature utilize only a small fraction of the available biomarker information. In particular, even though biomarkers are measured repeatedly over time, risk scores are typically based on the last available biomarker measurement. It is evident that such an approach discards valuable information because it does not take into account that the rate of change in the biomarker levels is not only different from patient to patient but also dynamically changes over time for the same patient. Hence, it is medically relevant to investigate whether repeated measurements of a biomarker can provide a better understanding of disease progression and a better prediction of the risk for the event of interest than a single biomarker measurement.

In line with the previous arguments, the motivation for this research comes from a study conducted by the Department of Cardio-Thoracic Surgery of the Erasmus Medical Center in the Netherlands. This study includes 285 patients who received a human tissue valve in the aortic position in the hospital from 1987 until 2008 \citep{bekkers.et.al:11}. Aortic allograft implantation has been widely used for a variety of aortic valve or aortic root diseases. Major advantages ascribed to allografts are the excellent hemodynamic characteristics as a valve substitute; the low rate of thrombo-embolic complications, and, therefore, absence of the need for anticoagulant treatment; and the resistance to endocarditis. A major disadvantage of using human tissue valves, however is the susceptibility to degeneration and the concomitant need for re-interventions. The durability of a cryopreserved aortic allograft is age-dependent, leading to a high lifetime risk of re-operation, especially for young patients. Re-operations on the aortic root are complex, with substantial operative risks, and mortality rates in the range 4--12\%. It is therefore of great interest for cardiologists and cardio-thoracic surgeons to have at their disposal an accurate prognostic tool that will inform them about the future prospect of a patient with a human tissue valve in order to optimize medical care, carefully plan re-operation and minimize valve-relate morbidity and mortality.

From the statistical analysis viewpoint the challenge is to utilize a technique capable of updating estimates of survival probabilities for a new patient as additional longitudinal information is recorded. An early approach in solving this problem has been landmarking \citep{anderson.et.al:83, zheng.heagerty:05, vanhouwelingen:07}. The basic idea behind landmarking is to obtain survival probabilities from a Cox model fitted to the patients from the original dataset who are still at risk at the time point of interest (e.g., the last time point we know that the new patient was still alive). A relatively newer method for producing dynamic predictions of survival probabilities is based on the class of joint models for longitudinal and time-to-event data \citep{henderson.et.al:02, yu.et.al:08, proust-lima.taylor:09, rizopoulos:11, rizopoulos:12}. In these models we have a complete specification of the joint distribution of the longitudinal response and the event times based on which the predictions in question can be derived. The main aim of this paper is to further study and contrast these two approaches. In particular, we show how survival probabilities are obtained under each method and what the differences are in the underlying assumptions. In addition, we focus on the functional relationship between the two processes and how this may affect predictions. We surpass the standard formulation, which only includes the current value of the marker, and we postulate functional forms that allow the rate of increase/decrease of the longitudinal outcome or a suitable summary of the whole longitudinal trajectory to determine the risk for an event. To assess the quality of the derived predictions from the two approaches we present different measures of discrimination and calibration, suitably adjusted to the context of longitudinal biomarkers.

The rest of the paper is organized as follows. Section~\ref{Sec:DynPred} describes formally the context of dynamic predictions and presents the landmarking and joint modeling approaches. Section~\ref{Sec:FunForm} shows different options for the functional form of the association structure between the longitudinal and event time processes. Section~\ref{Sec:PredAcc} presents measures of discrimination and calibration adapted to the dynamic predictions setting. Section~\ref{Sec:Appl} illustrates the use of joint modeling and landmarking in the Aortic Valve dataset and Section~\ref{Sec:Simul} refers to the results of a simulation study. Finally, Section~\ref{Sec:Disc} concludes the paper.


\section{Dynamic Individualized Predictions} \label{Sec:DynPred}
Following the discussion in Section~\ref{Sec:Intro} and the motivation from the Aortic Valve dataset, we present here the two frameworks for deriving dynamic individualized predictions. Let $\mathcal D_n = \{T_i, \delta_i, \by_i; i = 1, \ldots, n\}$ denote a sample from the target population, where $T_i^*$ denotes the true event time for the $i$-th subject ($i = 1, \ldots, n$), $C_i$ the censoring time, $T_i = \min(T_i^*, C_i)$ the corresponding observed event time, and $\delta_i = I(T_i^* \leq C_i)$ the event indicator, with $I(\cdot)$ being the indicator function that takes the value 1 when $T_i^* \leq C_i$, and 0 otherwise. In addition, we let $\by_i$ denote the $n_i \times 1$ longitudinal response vector for the $i$-th subject, with element $y_{il}$ denoting the value of the longitudinal outcome taken at time point $t_{il}$, $l = 1, \ldots, n_i$.

We are interested in deriving predictions for a new subject $j$ from the same population that has provided a set of longitudinal measurements $\mathcal Y_j(t) = \{ y_j(t_{jl}); 0 \leq t_{jl} \leq t, l = 1, \ldots, n_j \}$, and has a vector of baseline covariates $\bw_j$. The fact that biomarker measurements have been recorded up to $t$, implies survival of this subject up to this time point, meaning that it is more relevant to focus on the conditional subject-specific predictions, given survival up to $t$. In particular, for any time $u > t$ we are interested in the probability that this new subject $j$ will survive at least up to $u$, i.e.,
\[
\pi_j(u \mid t) = \Pr ( T_j^* \geq u \mid T_j^* > t, \mathcal Y_j(t), \bw_j, \mathcal D_n).
\]
The time-dynamic nature of $\pi_j(u \mid t)$ is evident because when new information is recorded for patient $j$ at time $t' > t$, we can update these predictions to obtain $\pi_j(u \mid t')$, and therefore proceed in a time-dynamic manner.


\subsection{Landmarking} \label{Sec:DynPred-Landmarking}
The landmarking approach provides an estimate of $\pi_j(u \mid t)$ by selecting the subjects at risk at $t$ from the original dataset $\mathcal D_n$, and using these to derive predictions. More formally, let $\mathcal R(t) = \{i: T_i > t\}$ denote the adjusted risk set, including all subjects who were not censored or dead by the landmark time $t$. Then a Cox model is fitted to these subjects by resetting time with zero being the landmark time, i.e.,
\begin{eqnarray*}
h_i(u-t) & = & \lim_{\Delta t \rightarrow 0} \frac{1}{\Delta t} \Pr \bigl \{ u - t \leq T_i^* < u - t + \Delta t \mid T_i^* > u - t, \mathcal Y_i(t) \bigr \}\\
& = & h_0(u-t) \exp \bigl \{ \bfgamma^\top \bw_i + \alpha \tilde y_i(t) \bigr \}, u > t,
\end{eqnarray*}
where the baseline hazard function $h_0(\cdot)$ is assumed completely unspecified, $\bw_i$ denotes a vector of baseline covariates, and the last available longitudinal response $\tilde y_i(t)$ also enters into the model as an ordinary baseline covariate. Having fitted this Cox model, an estimate of $\pi_j(u \mid t)$ is simply obtained by means of the Breslow estimator:
\begin{equation}
\hat \pi_j^{LM}(u \mid t) = \exp \Bigl [ - \widehat H_0(u) \exp \{ \widehat \bfgamma^\top \bw_j + \hat \alpha \tilde y_j(t) \} \Bigr ] \label{Eq:pi.u.t-LM},
\end{equation}
where
\[
\widehat H_0(u) = \sum \limits_{i \in \mathcal R(t)} \frac{I(T_i \leq u)
\delta_i}{ \sum_{\ell \in \mathcal R(u)} \exp \{\widehat \bfgamma^\top \bw_\ell + \hat \alpha \tilde y_\ell(t) \}}.
\]
\citet{vanhouwelingen:07} and \citet{zheng.heagerty:05} discuss several extensions of this approach that have greater flexibility by allowing the regression coefficient $\alpha$ to depend on time, i.e.,
\[
h_i(u-t) = h_0(u-t) \exp \bigl \{ \bfgamma^\top \bw_i + \alpha(u-t) \tilde y_i(t) \bigr \},
\]
and also, possibly, a baseline hazard that is not only a function of the time since the last measurement $u - t$, but also a function of the measurement time $t$, relaxing thus the proportional hazards assumption. An advantage of landmarking is that it can be very easily applied in practice, because it only requires fitting a simple Cox model each time a new measurement has been recorded for the subject for whom predictions are of interest.


\subsection{Joint Modeling} \label{Sec:DynPred-JM}
Contrary to the landmark approach, in the framework of joint models for longitudinal and time-to-event data we have a complete specification of the joint distribution of the two outcomes \citep{faucett.thomas:96, wulfsohn.tsiatis:97, henderson.et.al:00, tsiatis.davidian:04, rizopoulos:12}. For the longitudinal biomarker measurements mixed-effects models are typically employed to describe the subject-specific longitudinal trajectories. For simplicity of exposition and because the marker that we are going to use for the Aortic Valve dataset, namely the aortic gradient, is a continuous one, we focus here on linear mixed-effects models,
\begin{eqnarray}
\begin{array}{rcl}
y_i(t) & = & m_i(t) + \eps_i(t) = \bx_i^\top(t) \bfbeta + \bz_i^\top(t) \bb_i + \eps_i(t),\\
\bb_i & \sim & \mathcal N (\mathbf{0}, \bD), \quad \eps_i(t) \sim \mathcal N (0, \sigma^2),\\
\end{array} \label{Eq:LinearMixed}
\end{eqnarray}
where $y_i(t)$ denotes the observed value of the longitudinal outcome at any particular time point $t$, $\bx_i(t)$ and $\bz_i(t)$ denote the time-dependent design vectors for the fixed-effects $\bfbeta$ and for the random effects $\bb_i$, respectively, and $\eps_i(t)$ the corresponding error terms that are assumed independent of the random effects, and $\mbox{cov} \{\eps_i(t), \eps_i(t')\} = 0$ for $t' \neq t$. For the survival process, we assume that the risk for an event depends on the `true' and unobserved value of the marker at time $t$ (i.e., excluding the measurement error), denoted by $m_i(t)$ in \eqref{Eq:LinearMixed}. More specifically, we have
\begin{eqnarray}
\nonumber h_i (t \mid \mathcal M_i(t), \bw_i) & = & \lim_{\Delta t \rightarrow 0} \frac{1}{\Delta t}\Pr \{ t \leq T_i^* < t + \Delta t \mid T_i^* \geq t, \mathcal M_i(t), \bw_i \} \\
& = & h_0(t) \exp \bigl \{ \bfgamma^\top
\bw_i + \alpha m_i(t) \bigr \}, \quad t > 0, \label{Eq:Surv-RR}
\end{eqnarray}
where $\mathcal M_i(t) = \{ m_i(s), 0 \leq s < t \}$ denotes the history of the true unobserved longitudinal process up to $t$, $h_0(\cdot)$ denotes the baseline hazard function, and, as before, $\bw_i$ is a vector of baseline covariates with corresponding regression coefficients $\bfgamma$. Parameter $\alpha$ quantifies the association between the true value of the marker at $t$ and the hazard for an event at the same time point. Estimation of joint model's parameters can be based either on maximum likelihood or a Bayesian approach using Markov chain Monte Carlo algorithms. The likelihood of the model is derived under the assumptions that given the random effects, both the longitudinal and event time process are assumed independent, and the longitudinal responses of each subject are assumed independent. Formally we have,
\begin{eqnarray}
p(\by_i, T_i, \delta_i \mid \bb_i, \bftheta) & = & p(\by_i \mid \bb_i, \bftheta) \; p(T_i, \delta_i \mid \bb_i, \bftheta), \label{Eq:CondInd-I}\\
p(\by_i \mid \bb_i, \bftheta) & = & \prod_l p ( y_{il} \mid \bb_i, \bftheta ), \label{Eq:CondInd-II}
\end{eqnarray}
where $\bftheta^\top = (\bftheta_t^\top, \bftheta_y^\top, \bftheta_b^\top)$ denotes the full parameter vector, with $\bftheta_t$ denoting the parameters for the event time outcome, $\bftheta_y$ the parameters for the longitudinal outcomes, and $\bftheta_b$ the unique parameters of the random-effects covariance matrix, and $p(\cdot)$ denotes an appropriate probability density function. More details regarding the estimation and properties of joint models can be found in \citet{rizopoulos:12} and \citet[Chapter~7]{ibrahim.et.al:01}.

Under this framework, estimation of $\pi_j(u \mid t)$ can be based on (asymptotic) Bayesian arguments and the corresponding posterior predictive distribution:
\[
\pi_j(u \mid t) = \int \Pr(T_j^* \geq u \mid T_j^* > t, \mathcal Y_j(t), \bftheta) \, p(\bftheta \mid \mathcal D_n) \, d\bftheta.
\]
The calculation of the first part of each integrand takes full advantage of the conditional independence assumptions \eqref{Eq:CondInd-I} and \eqref{Eq:CondInd-II}. In particular, we observe that the first term of the integrand of $\pi_j(u \mid t)$ can be rewritten by noting that:
\begin{eqnarray*}
\nonumber \Pr(T_j^* \geq u \mid T_j^* > t, \mathcal Y_j(t), \bftheta) & = & \int \Pr(T_j^* \geq u \mid T_j^* > t, \bb_j, \bftheta) \, p(\bb_j \mid T_j^* > t, \mathcal Y_j(t), \bftheta) \, d\bb_j\\
& = & \int \frac{S_j \bigl \{ u \mid \mathcal M_j(u, \bb_j), \bftheta \bigr \}}{S_j \bigl \{ t \mid \mathcal M_j(t, \bb_j), \bftheta \bigr \} } \, p(\bb_j \mid T_j^* > t, \mathcal Y_j(t), \bftheta) \, d\bb_j,
\end{eqnarray*}
where
\[
S_j \bigl \{ t \mid \mathcal M_j(t, \bb_j), \bftheta \bigr \} = \exp \biggl \{ \int_0^t h_0(s) \exp\{\bfgamma^\top
\bw_i + \alpha m_i(s) \} \biggr \} \; ds,
\]
denotes the subject-specific survival function.

Combining these equations with the maximum likelihood estimates or with the MCMC sample from the posterior distribution of the parameters for the original data $\mathcal D_n$, we can devise a simple simulation scheme to obtain a Monte Carlo estimate of $\pi_j(u \mid t)$. More specifically, this is comprised of the following steps:
\begin{itemize}
\item[Step 1.] Take $K$ samples of $\{\bftheta^{(k)}, k = 1, \ldots, K\}$ from either the MCMC sample of $p(\bftheta \mid \mathcal D_n)$ or the asymptotic normal posterior distribution $\mathcal N (\widehat \bftheta, \mathcal H_n)$, where $\widehat \bftheta$ denotes the maximum likelihood estimates and $\mathcal H_n$ the observed information matrix
    \[
    \mathcal H_n = \biggl \{ -\sum \limits_{i = 1}^n \frac{\partial^2 \log p(\by_i, T_i, \delta_i, \bftheta)}{\partial \bftheta^\top \partial \bftheta} \Big |_{\theta = \hat \theta} \biggr \}^{-1}.
    \]

\item[Step 2.] Draw $K$ realizations $\{\bb_j^{(k)}, k = 1, \ldots, K\}$ for the random effects of the new subject $j$ from the posterior distribution of the random effects \[
    p \bigl ( \bb_j \mid T_j^* > t, \mathcal Y_j(t), \bftheta^{(k)} \bigr ) \propto \biggl \{ \prod \limits_{l = 1}^{n_j(t)} p \bigl ( y_{jl} \mid \bb_j, \bftheta^{(k)} \bigr ) \biggr \} S_j \bigl \{ t \mid \mathcal M_j(t, \bb_j), \bftheta^{(k)} \bigr \} p \bigl ( \bb_j, \bftheta^{(k)} \bigr ),
    \]
    where $n_j(t)$ denotes the number of available measurements for subject $j$ by time $t$.
\item[Step 3.] Based on these realizations an estimate of $\pi_j(u \mid t)$ is derived as
    \begin{equation}
    \hat \pi_j^{JM} (u \mid t) = \frac{1}{K} \sum \limits_{k = 1}^K \frac{S_j \bigl \{ u \mid \mathcal M_j(u, \bb_j^{(k)}), \bftheta^{(k)} \bigr \}}{S_j \bigl \{ t \mid \mathcal M_j(t, \bb_j^{(k)}), \bftheta^{(k)} \bigr \}}. \label{Eq:pi.u.t-JM}
    \end{equation}
\end{itemize}
More details can be found in \citet{yu.et.al:08} and \citet{rizopoulos:11, rizopoulos:12}.


\subsection{Heuristic Comparison between Landmarking and Joint Modeling} \label{Sec:DynPred-Comp}
The previous two sections illustrated that both landmarking and joint modeling can be utilized to derive dynamically updated estimates of conditional survival probabilities $\pi_j(u \mid t)$. The landmark approach can be more easily implemented in practice because it only requires fitting a standard Cox model, whereas joint models require specialized software \citep{rizopoulos:10, rizopoulos:12}. In addition, joint models seem to make more modeling assumptions than the landmark approach, which poses a concern regarding how a misspecification of these assumptions may affect predictions. On the other hand, the landmark approach uses less information than joint modeling (i.e., only the last observed longitudinal response), and hence is less optimal. The following points provide a more detailed exposition of the underlying differences between the two approaches.
\begin{itemize} \itemsep=1cm
\item \textbf{Extrapolation:} The main differences in how landmarking and joint modeling tackle the problem of prediction can best be explained by Figure~\ref{Fig:LMvsJM}. This shows the longitudinal responses of a hypothetical subject who was alive up to year five and for whom we would like to obtain a predicted survival function. To produce estimates of the conditional survival probabilities both landmarking and joint modeling require a value for the longitudinal response at $t = 5$ (vertical dotted line). Since this subject provided her last longitudinal measurement at year three, some sort of extrapolation is taking place. In particular, landmarking is based on a `last value carried forward' approach and uses as the value of the longitudinal response at year five the last available measurement of the subject at year three (horizontal dashed line). Even though this approach is conceptually simple and easy to perform in practice, unfortunately, it may lead to biased and misleading inference on the Cox model parameters \citep{tsiatis.davidian:01}. Joint modeling on the other hand uses the subject-specific fitted value of the longitudinal profile from the linear mixed model extrapolated at year 5, i.e., $m_j(5) = \bx_j^\top(5) \bfbeta + \bz_j^\top(5) \bb_j$ (solid line). This approach uses all available information, because the estimate of $m_j(5)$ is based on both all past values of this subject and on the responses of other subjects. To explain how the borrowing of information between subjects is taking place, assume, hypothetically, that there was another patient, who during the first three years had exactly the same longitudinal measurements as the patient depicted in Figure~\ref{Fig:LMvsJM}, but also she had extra measurements up to year five. The joint model would make use of this patient and say that the profile of the patient in Figure~\ref{Fig:LMvsJM} would be similar to the one of the patient with the extra measurements. From a biological point of view the joint modeling approach seems more logical than landmarking because we indeed expect the biomarker levels of a patient to continuously change over time rather than to remain constant between visits.

    Note that in general even if we had observed the longitudinal response at $t = 5$, i.e., $y_j(5)$ this will not be equal to $m_j(5)$. The joint model assumes that the realizations of the longitudinal marker are the output of a stochastic process generated by the subject, and it is the underlying signal in the process, represented by $m_j(t)$, that is associated with the hazard for an event. The observed data $y_j(t)$ are a contaminated with measurement error version of the underlying signal $m_j(t)$. This measurement error most often stems from biological variation, but some times may also be attributed to the medical test/examination used to measure the marker.
    \begin{figure}[!h]
    \includegraphics[width = \textwidth]{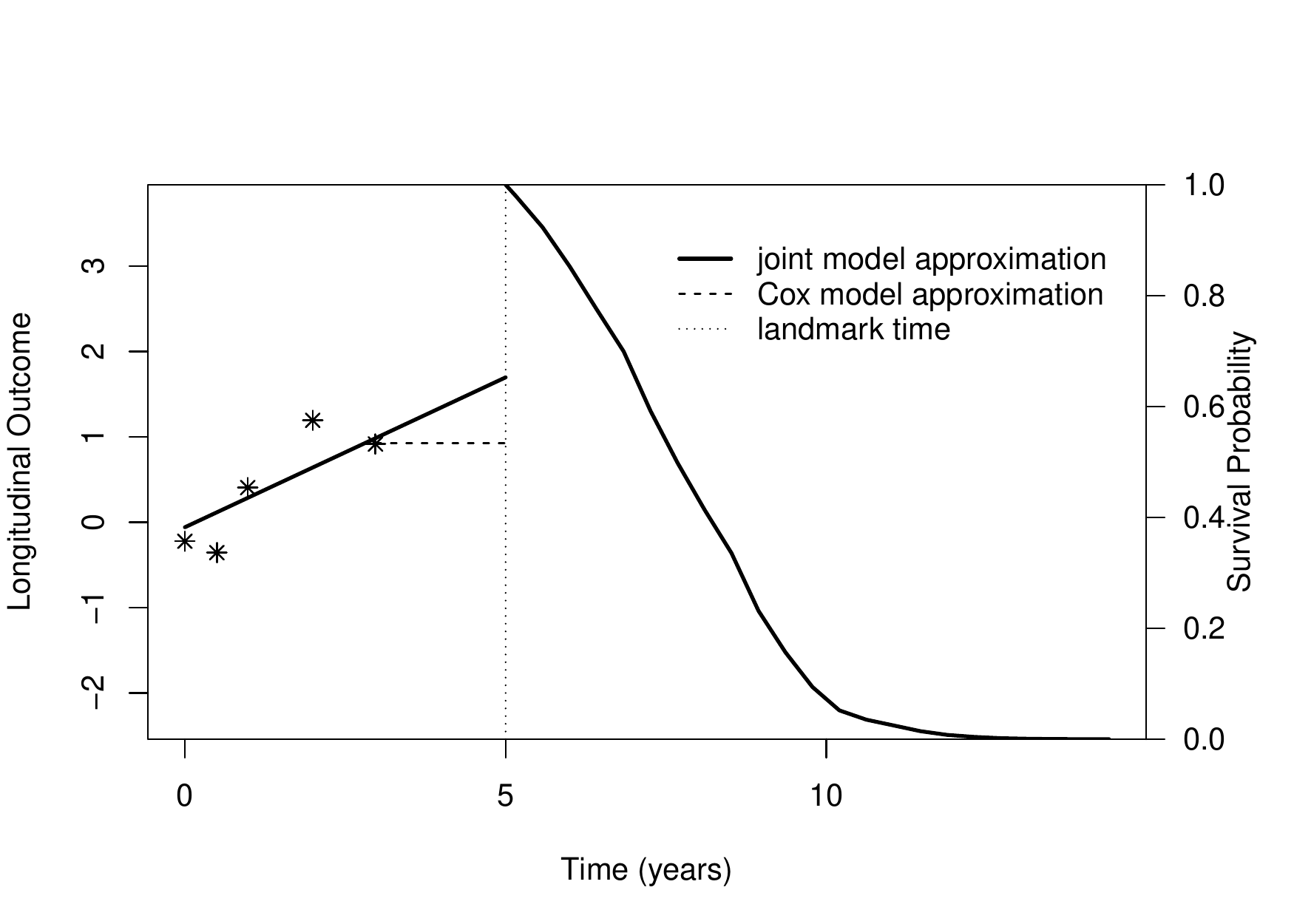}
    \caption{Graphical comparison on how landmarking and joint modeling use the available longitudinal measurements to provide an estimate of the longitudinal outcome at the last time point the patients was still alive. The left side of the plot shows the observed longitudinal responses, and the fitted longitudinal profile from the joint model. The right side shows the corresponding survival probability.} \label{Fig:LMvsJM}
    \end{figure}

\item \textbf{Assumptions related to the longitudinal process:} The landmark approach assumes that the visiting process, which is the process, stochastic or deterministic, that generates the visit times at which subjects provide measurements is independent of the longitudinal marker process and the survival time $T_j^*$. The joint modeling approach also assumes that a visit scheduled at time $t$ is independent of a future event occurring at $T_j^* > t$ and of future longitudinal responses $\{y_j(s), t \leq s \leq T_j^*\}$, but it does allow visit times to depend on the observed longitudinal responses $\mathcal Y_j(t)$. This is a more realistic assumption because what we expect to happen in practice is that physicians will ask a patient to come back more often if they observe a worsening of her condition based on her observed responses. In addition, subjects may have missing marker measurements during follow-up. The landmark approach assumes that any such missingness is completely at random \citep{little.rubin:02}. On the other hand, due to the fact that joint modeling is based on a complete specification of the joint likelihood function of the longitudinal and event time processes, it allows incomplete longitudinal data to be missing at random. Hence, joint modeling is capable of providing valid inferences under less stringent assumptions than landmarking. Though, it should be mentioned that these advantageous features require the joint model to be roughly correctly specified.

\item \textbf{Assumptions related to the event process:} Similarly to the assumptions for the longitudinal process, landmarking makes more stringent assumptions for the censoring process. In particular, under the landmark approach censoring is assumed independent of past longitudinal responses $\{y_j(s); 0 \leq s < t \}$, whereas under joint modeling and again because we use a complete specification of the joint likelihood function, censoring is allowed to depend in a general way on $\{y_j(s); 0 \leq s < t \}$.
\end{itemize}


\section{Functional Form} \label{Sec:FunForm}
The assessment of the predictive value of baseline covariates is to a degree simple, in the sense that these covariates are typically included in a prognostic model as is or under a suitable transformation (e.g., log-scale, polynomials, splines, etc.). However, in our setting, where we have multiple longitudinal measurements available per subject there could be different features of the longitudinal process that are most predictive for the event of interest. For example, in ordinary proportional hazards models, it has been long recognized that the functional form of time-varying covariates influences the derived inferences; see, for instance, \citet{fisher.lin:99} and references therein. In the joint modeling framework however, where the longitudinal outcome plays the role of a time-dependent covariate for the survival process, this topic has received less attention. The two main functional forms that have been primarily used so far in joint models include in the linear predictor of the relative risk model \eqref{Eq:Surv-RR} either the subject-specific means $m_i(t)$ from the longitudinal submodel or just the random effects $\bb_i$ \citep{henderson.et.al:00, rizopoulos.ghosh:11}. However, as argued above, there could be other characteristics of the patients' longitudinal profiles that are more predictive for the risk of an event, such as the rate of increase/decrease of the biomarker's levels or a suitable summary of the whole longitudinal trajectory. Here we present a few examples of alternative formulations for the association structure between the longitudinal outcome and the risk for an event:
\begin{eqnarray}
h_i(t) & = & h_0(t) \exp \bigl \{ \bfgamma^\top \bw_i + \alpha_1 m_i(t)  + \alpha_2 m_i'(t) \bigr \}, \quad m_i'(t) = \frac{d m_i(t)}{dt},\label{Eq:Param-TDslopes}\\
h_i(t) & = & h_0(t) \exp \Bigl \{ \bfgamma^\top \bw_i + \alpha \int_0^t m_i(s) \, ds \Bigr \}, \label{Eq:Param-CumEff}\\
h_i(t) & = & h_0(t) \exp \Bigl \{ \bfgamma^\top \bw_i + \alpha \int_0^t \varrho(t - s) m_i(s) \, ds \Bigr \}, \label{Eq:Param-wCumEff}\\
h_i(t) & = & h_0(t) \exp ( \bfgamma^\top \bw_i + \bfalpha^\top \bb_i ). \label{Eq:Param-RE}
\end{eqnarray}
It is evident that these parameterizations have different sets of association parameters $\bfalpha$, and in addition that the interpretation of these parameters is different for each formulation. In particular, parameterization \eqref{Eq:Param-TDslopes} postulates that the risk for an event at a particular time point $t$ depends not only on the level of the marker at this time point but also on its rate of change, captured by the slope term $m_i'(t)$. This could be of importance when two patients at a specific time point have equal marker levels, but one patient having an increasing trajectory and the other a decreasing one. Parameterization \eqref{Eq:Param-CumEff} posits that the risk for an event at time $t$ is associated with the area under the longitudinal trajectory up to this point. This can be considered as a summary of the whole marker history up to $t$ and contrary to the previous formulations it allows the risk to the depend on the whole history $\mathcal M_i(t) = \{ m_i(s), 0 \leq s < t \}$ and not only on features of the marker at $t$. Parameterization \eqref{Eq:Param-wCumEff} extends \eqref{Eq:Param-CumEff} by assigning to the past values of the longitudinal trajectory different weights, using a function $\varrho(\cdot)$. For instance, setting $\varrho(t - s) = \phi(t - s) / \{ \Phi(t) - 0.5 \}$, where $0 < s < t$, and $\phi(\cdot)$ and $\Phi(\cdot)$ denote the probability density and cumulative distribution functions of the standard normal distribution, respectively, we assume that the risk at $t$ only depends on the marker levels in the interval $(t - 3, t)$ with values closer to $t$ having higher weight, because when $t - s > 3$ then $\varrho(t - s)$ is practically zero. Finally, parameterization \eqref{Eq:Param-RE} is time-independent and assumes that the hazard for an event is related to the random effects from the longitudinal process. This formulation shares similarities with the time-dependent slopes parameterization \eqref{Eq:Param-TDslopes} when a simple random-intercepts and random-slopes structure is assumed for the longitudinal submodel.

Under the landmarking approach, and in order to improve predictive performance, we could also make better use of the observed longitudinal history than just using the last available measurement. Mimicking the formulations presented above for joint modeling, we can define Cox models fitted to the patients at risk at the landmark time $t$, which include $\tilde y_i'(t)$ that denotes the slope calculated from the last two available measurements of each subject, and $\sum_{0 \leq s \leq t} y_i(s) \Delta s$ that denotes the area under the step function defined from the observed longitudinal measurements up to $t$:
\begin{eqnarray*}
h_i(u - t) & = & h_0(u - t) \exp \bigl \{ \bfgamma^\top \bw_i + \alpha_1 \tilde y_i(t) + \alpha_2 \tilde y_i'(t) \bigr \}, \label{Eq:Param-TDslopes-LM}\\
h_i(u - t) & = & h_0(u - t) \exp \Bigl \{ \bfgamma^\top \bw_i + \alpha \sum \limits_{s=0}^t y_i(s) \Delta s \Bigr \}, \label{Eq:Param-CumEff-LM}\\
h_i(u - t) & = & h_0(u - t) \exp \Bigl \{ \bfgamma^\top \bw_i + \alpha \sum \limits_{s=0}^t \varrho(t - s) y_i(s) \Delta s \Bigr \}, \label{Eq:Param-wCumEff-LM}
\end{eqnarray*}
where, as before, $\varrho(t - s)$ is a potential weight function. Note that we do not have an analogous functional form to \eqref{Eq:Param-RE} under landmarking.


\section{Measuring Predictive Performance} \label{Sec:PredAcc}
The assessment of the predictive performance of time-to-event models has received a lot of attention in the statistical literature. In general, the developed methodology has focused on calibration, i.e., how well the model predicts the observed data \citep{schemper.henderson:00, gerds.schumacher:06} or discrimination, i.e., how well can the model discriminate between patients that had the event from patients that did not \citep{harrell.et.al:96, pencina.et.al:08}. In the following we present discrimination and calibration measures suitably adapted to the dynamic prediction setting. It should be noted that these measures require in their essence an estimate of $\pi_j(u \mid t)$, and therefore they are applicable under both landmarking and joint modeling. In the following we will use the term $\hat \pi_j(u \mid t)$ to generically denote either \eqref{Eq:pi.u.t-LM} or \eqref{Eq:pi.u.t-JM}.

\subsection{Discrimination}
To take into account the dynamic nature of the longitudinal marker in discriminating between subjects, we focus on a time interval of medical relevance within which the occurrence of events is of interest. In this setting, a useful property of the model would be to successfully discriminate between patients who are going to experience the event within this time frame from patients who will not. To put this formally, as before, we assume that we have collected longitudinal measurements $\mathcal Y_j(t) = \{ y_j(t_{jl}); 0 \leq t_{jl} \leq t, l = 1, \ldots, n_j \}$ up to time point $t$ for subject $j$. We are interested in events occurring in the medically-relevant time frame $(t, t + \Delta t]$ within which the physician can take an action to improve the survival chance of the patient. Under the assumed model and the methodology presented in Section~\ref{Sec:DynPred}, we can define a prediction rule using $\pi_j(t + \Delta t \mid t)$ that takes into account the available longitudinal measurements $\mathcal Y_j(t)$. In particular, for any value $c$ in $[0, 1]$ we can term subject $j$ as a case if $\pi_j(t + \Delta t \mid t) \leq c$ (i.e., occurrence of the event) and analogously as a control if $\pi_j(t + \Delta t \mid t) > c$. Thus, in this context, we define sensitivity and specificity as
\[
\Pr \bigl \{ \pi_j(t + \Delta t \mid t) \leq c \mid T_j^* \in (t, t + \Delta t] \bigr \},
\]
and
\[
\Pr \bigl \{ \pi_j(t + \Delta t \mid t) > c \mid T_j^* > t + \Delta t \bigr \},
\]
respectively. For a randomly chosen pair of subjects $\{i, j\}$, in which both subjects have provided measurements up to time $t$, the discriminative capability of the assumed model can be assessed by the area under the receiver operating characteristic curve (AUC), which is obtained for varying $c$ and equals,
\[
\mbox{AUC}(t, \Delta t) = \Pr \bigl [ \pi_i(t + \Delta t \mid t) < \pi_j(t + \Delta t \mid t) \mid \{ T_i^* \in (t, t + \Delta t] \} \cap \{ T_j^* > t + \Delta t \} \bigr ], \label{Eq:AUCt}
\]
that is, if subject $i$ experiences the event within the relevant time frame whereas subject $j$ does not, then we would expect the assumed model to assign higher probability of surviving longer than $t + \Delta t$ for the subject who did not experience the event. To summarize the discriminative power of the assumed model over the whole follow-up period, we need to take into account that the number of subjects contributing to the comparison of the fitted $\pi_i(t + \Delta t \mid t)$ with the observed data is not the same for all time points $t$. Following an approach similar to \citet{antolini.et.al:05} and \citet{heagerty.zheng:05}, we propose the use of a weighted average of AUCs
\begin{equation}
\mbox{C}_{dyn}^{\Delta t} = \int_0^\infty \mbox{AUC}(t, \Delta t) \, \Pr \{ \mathcal E(t) \} \; dt \Big / \int_0^\infty \Pr \{ \mathcal E(t) \} \; dt, \label{Eq:dynC}
\end{equation}
where $\mathcal E(t) = \bigl [ \{ T_i^* \in (t, t + \Delta t] \} \cap \{ T_j^* > t + \Delta t \} \bigr ]$, and $\Pr \{ \mathcal E(t) \}$ denotes the probability that a random pair is comparable at $t$. We call $\mbox{C}_{dyn}^{\Delta t}$ the dynamic concordance index since it summarizes the concordance probabilities over the follow-up period. Note also that $\mbox{C}_{dyn}^{\Delta t}$ depends on the length $\Delta t$ of the time interval of interest, which implies that different models may exhibit different discrimination power for different $\Delta t$.

For the estimation of $\mbox{C}_{dyn}^{\Delta t}$ we need to take care of two issues, namely, the calculation of the integrals in the definition of \eqref{Eq:dynC} and censoring. For the former we use a 15-point Gauss-Kronrod quadrature rule \citep{press.et.al:07}. To take into account the fact that the number of subjects decreases over time due to the occurrence of events and censoring, for any time point $t$ we define as comparable pairs the pairs that satisfy the relation
\begin{eqnarray*}
\Omega_{ij}(t) & = & \bigl [\{T_i \in (t, t + \Delta t] \} \cap \{\delta_i = 1 \} \bigr ] \cap \{ T_j > t + \Delta t \} \mbox{ or }\\
&& \bigl [ \{T_i \in (t, t + \Delta t]\} \cap \{ \delta_i = 1 \} \bigr ] \cap \bigl [ \{T_j = t + \Delta t \} \cap \{\delta_j = 0\} \bigr ],
\end{eqnarray*}
where $i, j = 1, \ldots, n$ with $i \neq j$. For two comparable subjects $i$ and $j$, we can estimate and compare their survival probabilities $\pi_i(t + \Delta t \mid t)$ and $\pi_j(t + \Delta t \mid t)$, based on the methodology presented in Section~\ref{Sec:DynPred}. This leads to a natural estimator for $\mbox{AUC}(t, \Delta t)$ as the proportion of concordant subjects out of the set of comparable subjects for time $t$:
\[
\mbox{\aucHat}(t, \Delta t) = \frac{\sum_{i = 1}^n \sum_{j = 1; j \neq i}^n I \{ \hat \pi_i(t + \Delta t \mid t) < \hat \pi_j(t + \Delta t \mid t) \} \times I \{\Omega_{ij}(t)\}}{\sum_{i=1}^n \sum_{j=1; j \neq i}^n I\{\Omega_{ij}(t)\}},
\]
where $I(\cdot)$ denotes the indicator function. Having estimated $\mbox{AUC}(t, \Delta t)$, the next step in estimating $\mbox{C}_{dyn}^{\Delta t}$ is to obtain estimates for the weights $\mbox{Pr} \{ \mathcal E(t) \}$. We observe that these can be rewritten as
\begin{eqnarray*}
\Pr \{ \mathcal E(t) \} & = & \Pr \bigl [ \{ T_i^* \in (t, t + \Delta t] \} \cap \{ T_j^* > t + \Delta t \} \bigr ] \\
& = & \Pr (T_i^* \in (t, t + \Delta t]) \times \Pr( T_j^* > t + \Delta t)\\
& = & \{ S(t) - S(t + \Delta t) \} S(t + \Delta t),
\end{eqnarray*}
where the simplification in the second line comes from the independence of subjects $i$ and $j$, and $S(\cdot)$ here denotes the marginal survival function.

In practice the calculation of $\mbox{C}_{dyn}^{\Delta t}$ is restricted into a follow-up interval $[0, t_{max}]$ where we have information. Let $t_1, \ldots, t_{15}$ denote the re-scaled abscissas of the Gauss-Kronrod rule in the interval $[0, t_{max}]$ with corresponding weights $\varpi_1, \ldots, \varpi_{15}$. We combine the estimates $\mbox{\aucHat}(t_k, \Delta t)$, $k = 1, \ldots, 15$ with the estimates of the weights $\Pr \{ \mathcal E(t) \}$ to obtain
\[
\widehat{\mbox{C}}_{dyn}^{\Delta t} = \frac{\sum_{k = 1}^{15} \varpi_k \mbox{\aucHat}(t_k, \Delta t) \times \widehat{\mbox{Pr}} \{ \mathcal E(t_k) \}}{\sum_{k = 1}^{15} \varpi_k \widehat{\mbox{Pr}} \{ \mathcal E(t_k) \}},
\]
where $\widehat{\Pr} \{ \mathcal E(t_k) \} = \{\widehat S(t_k) - \widehat S(t_k + \Delta t)\} \widehat S (t_k + \Delta t)$, with $\widehat S (\cdot)$ denoting the Kaplan-Meier estimate of the marginal survival function $S(\cdot)$.


\subsection{Calibration}
The assessment of the accuracy of predictions of survival models is typically based on the expected error of predicting future events. In our setting, and again taking into account the dynamic nature of the longitudinal outcome, it is of interest to predict the occurrence of events at $u > t$ given the information we have recorded up to time $t$. This gives rise to expected prediction error:
\[
\mbox{PE}(u \mid t) = E \bigl [ L\{N_i(u) - \pi_i(u \mid t)\} \bigr ],
\]
where $N_i(t) = I(T_i^* > t)$ is the event status at time $t$, $L(\cdot)$ denotes a loss function, such as the absolute or square loss, and the expectation is taken with respect to the distribution of the event times. An estimate of $\mbox{PE}(u \mid t)$ that accounts for censoring has been proposed by \citet{henderson.et.al:02}:
\begin{eqnarray*}
\lefteqn{\nonumber \widehat{\mbox{PE}}(u \mid t) = \{n(t)\}^{-1} \sum_{i: T_i \geq t} I(T_i \geq u) L\{1 - \hat \pi_i(u \mid t)\} + \delta_i I(T_i < u) L\{0 - \hat \pi_i(u \mid t)\}}\\
&& + (1 - \delta_i) I(T_i < u) \Bigl [ \hat \pi_i(u \mid T_i) L\{1 - \hat \pi_i(u \mid t)\} + \{1 - \hat \pi_i(u \mid T_i)\} L\{0 - \hat \pi_i(u \mid t)\} \Bigr ],
\end{eqnarray*}
where $n(t)$ denotes the number of subjects at risk at time $t$. The first two terms in the sum correspond to patients who were alive after time $u$ and dead before $u$, respectively; the third term corresponds to patients who were censored in the interval $[t, u]$. Using the longitudinal information up to time $t$, $\mbox{PE}(u \mid t)$ measures the predictive accuracy at the specific time point $u$. Alternatively, we could summarize the error of prediction in a specific interval of interest, say $[t, u]$, by calculating a weighted average of $\{\mbox{PE}(s \mid t), t < s< u\}$ that corrects for censoring, similarly to $\mbox{C}_{dyn}^{\Delta t}$. An estimator of this type for the integrated prediction error has been suggested by \citet{schemper.henderson:00}, which adapted to our time-dynamic setting takes the form
\[
\mbox{\ipeHat}(u \mid t) = \frac{\sum_{i: t \leq T_i \leq u} \delta_i \bigl \{ \widehat S_C(t) / \widehat S_C(T_i) \bigr \} \widehat{\mbox{PE}}(u \mid t)}{\sum_{i: t \leq T_i \leq u} \delta_i \bigl \{ \widehat S_C(t) / \widehat S_C(T_i) \bigr \}},
\]
where $\widehat S_C(\cdot)$ denotes the Kaplan-Meier estimator of the censoring time distribution.

Both $\mbox{\ipeHat}(u \mid t)$ and $\widehat{\mbox{PE}}(u \mid t)$ can be used to provide a measure of explained variation between nested models. Assuming model $M_1$ is nested in model $M_2$, we can compute how much the extra structure in $M_2$ improves accuracy by
\[
R_{PE}^2(u \mid t; M_1, M_2) = 1 - \widehat{\mbox{PE}}_{M_2}(u \mid t) \Big / \widehat{\mbox{PE}}_{M_1}(u \mid t)
\]
or
\[
R_{IPE}^2(u \mid t; M_1, M_2) = 1 - \mbox{\ipeHat}_{M_2}(u \mid t) \Big / \mbox{\ipeHat}_{M_1}(u \mid t).
\]


\section{Analysis of the Aortic Valve Dataset} \label{Sec:Appl}
\begin{figure}[!h]
\includegraphics[width = \textwidth]{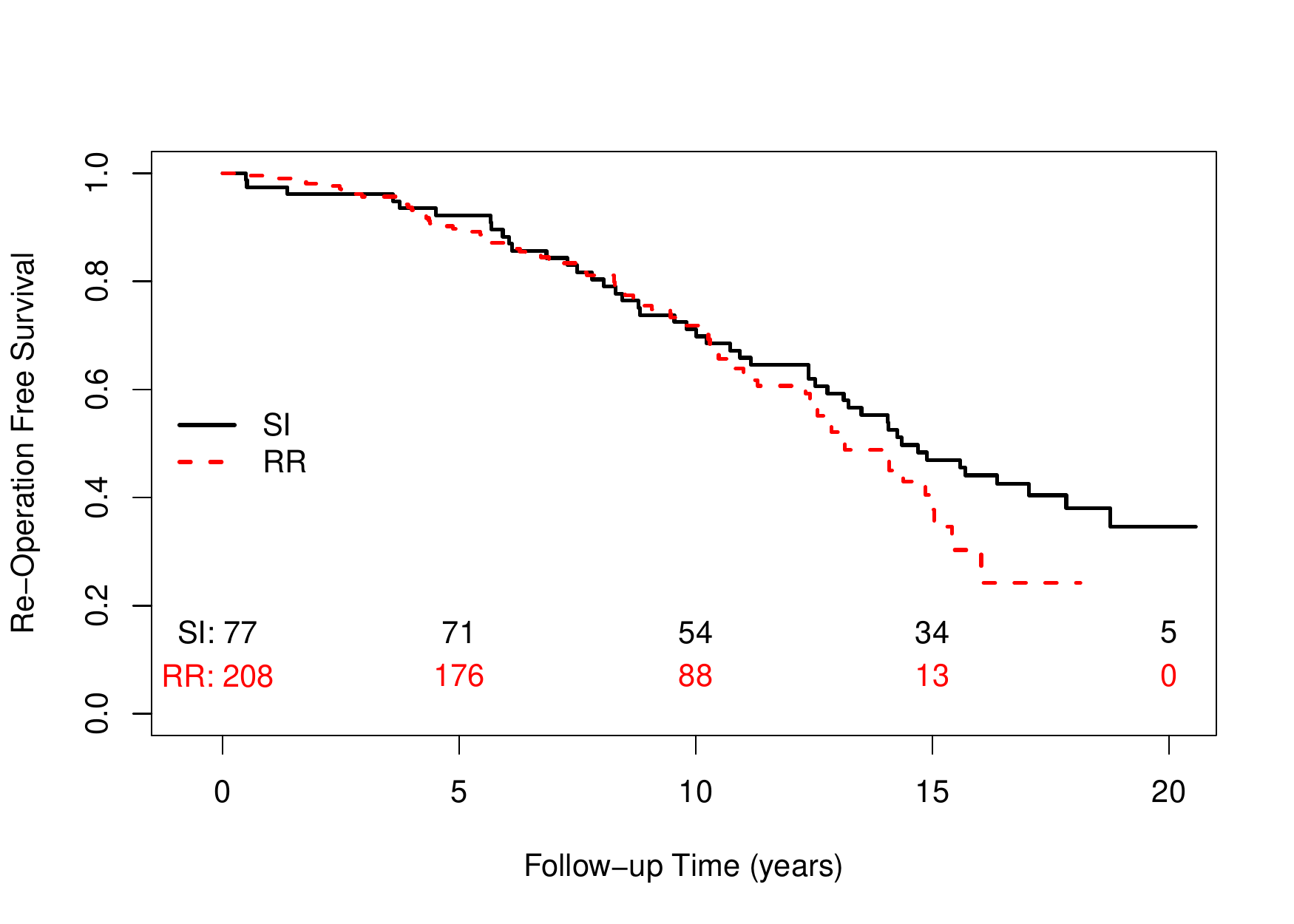}
\caption{Kaplan-Meier estimates of the survival functions for re-operation-free survival for the sub-coronary implantation (SI) and root replacement (RR) groups.} \label{Fig:KM}
\end{figure}

\begin{figure}[!h]
\includegraphics[width = \textwidth]{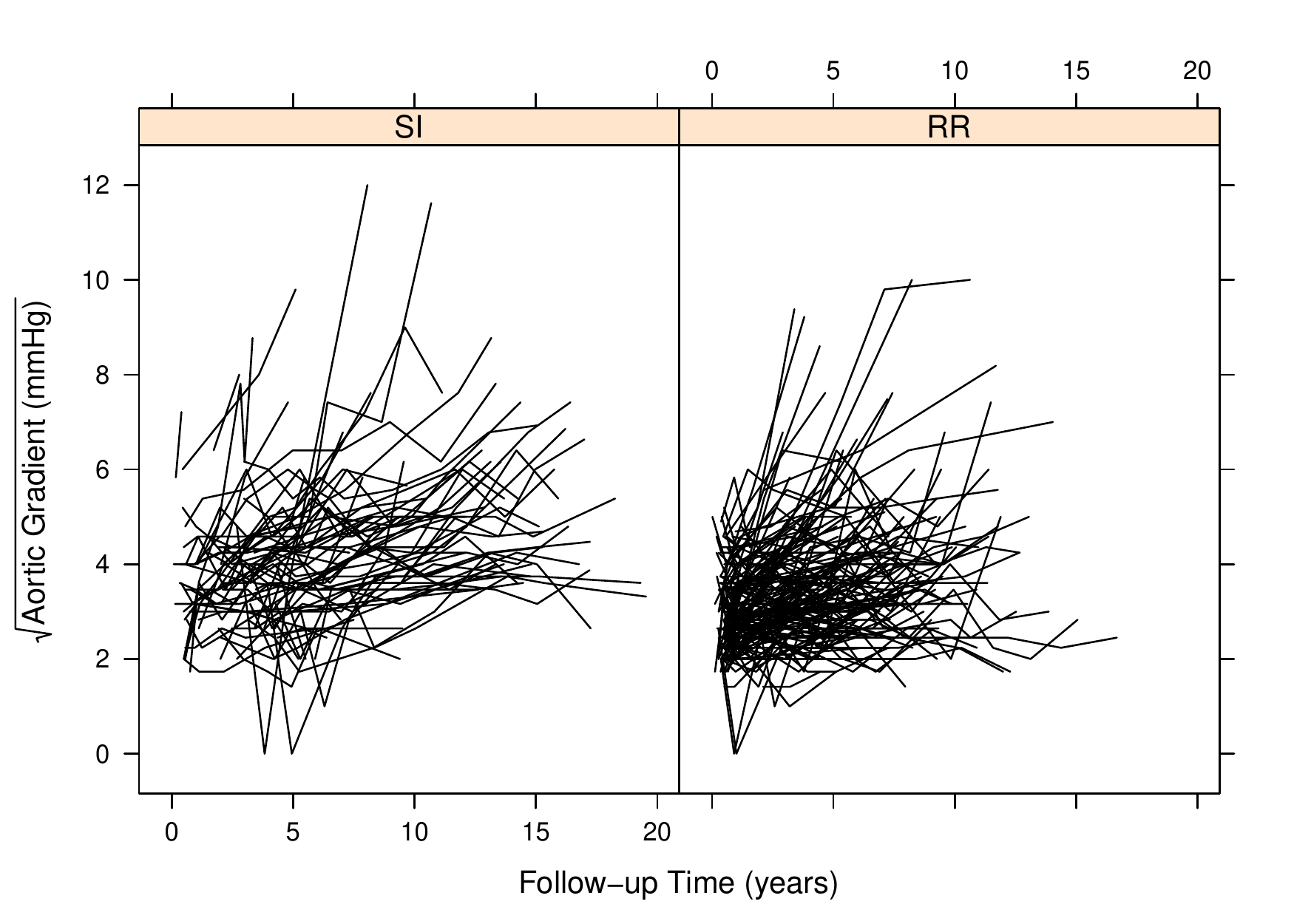}
\caption{Subject-specific profiles for the square root aortic gradient separately for the sub-coronary implantation (SI) and root replacement (RR) groups.} \label{Fig:SubjProfiles}
\end{figure}

We return to the Aortic Valve dataset introduced in Section~\ref{Sec:Intro}. Our aim is to use the existing data and provide accurate predictions of re-operation-free survival for future patients from the same population, utilizing their baseline information, namely age, gender, BMI and the type of operation they underwent, and their recorded aortic gradient levels. In our study, a total of 77 (27\%) patients received a sub-coronary implantation (SI) and the remaining 208 patients a root replacement (RR). These patients were followed prospectively over time with annual telephone interviews and biennial standardized echocardiographic assessment of valve function until July 8, 2010. Echo examinations were scheduled at 6 months and 1 year postoperatively, and biennially thereafter, and at each examination, echocardiographic measurements of aortic gradient (mmHg) were taken. By the end of follow-up, 1262 aortic gradient measurements were recorded with an average of 4.3 measurements per patient (s.d. 2.4 measurements), 59 (20.7\%) patients had died, and 73 (25.6\%) patients required a re-operation on the allograft. The composite event, re-operation or death, was observed for 125 (43.9\%) patients, and the corresponding Kaplan-Meier estimator for the two intervention groups is shown in Figure~\ref{Fig:KM}.
We can observe minimal differences in the re-operation-free survival rates between sub-coronary implantation and root replacement, with only a slight advantage of sub-coronary implantation towards the end of the follow-up. For the longitudinal process and because aortic gradient exhibits right skewness, we will proceed in our analysis using the square root transform of this outcome. Figure~\ref{Fig:SubjProfiles} depicts the subject-specific longitudinal profiles of the square root aortic gradient for the two intervention groups.
We observe considerable variability in the shapes of these trajectories, but there are no systematic differences apparent between the two groups.

We start by defining a set of joint models based on which predictions will be calculated. For the longitudinal process we allow a flexible specification of the subject-specific square root aortic gradient trajectories using natural cubic splines of time. More specifically, the linear mixed model takes the form
\begin{eqnarray*}
y_i(t) & = & \beta_1 \mbox{\tt SI}_i + \beta_2 \mbox{\tt RR}_i + \beta_3 \{B_1(t, \lambda) \times \mbox{\tt SI}_i\} + \beta_4 \{B_1(t, \lambda) \times \mbox{\tt RR}_i\}\\
&& \hspace{0.5cm} + \, \beta_5 \{B_2(t, \lambda) \times \mbox{\tt SI}_i\} + \beta_6 \{B_2(t, \lambda) \times \mbox{\tt RR}_i\}\\
&& \hspace{0.5cm} + \, \beta_7 \{B_3(t, \lambda) \times \mbox{\tt SI}_i\} + \beta_8 \{B_3(t, \lambda) \times \mbox{\tt RR}_i\}\\
&& \hspace{0.5cm} + \, b_{i0} + b_{i1}B_1(t, \lambda) + b_{i2}B_2(t, \lambda) + b_{i3}B_3(t, \lambda) + \eps_i(t),
\end{eqnarray*}
where $B_n(t, \lambda)$ denotes the B-spline basis for a natural cubic spline with boundary knots at baseline and 19 years and internal knots at 2.1 and 5.5 years (i.e., the 33.3\% and 66.6\% percentiles of the observed follow-up times), {\tt SI}  and {\tt RR} are the dummy variables for the sub-coronary implantation and root replacement groups, respectively, $\eps_i(t) \sim \mathcal N (0, \sigma^2)$ and $\bb_i \sim \mathcal N (\mathbf 0, \bD)$. For the survival process we consider four relative risk models, each positing a different association structure between the two processes, namely:
\begin{eqnarray*}
M_1: && h_i(t) = h_0(t) \exp \bigl \{\gamma_1 \mbox{\tt RR}_i + \gamma_2 \mbox{\tt Age}_i + \gamma_3 \mbox{\tt Female}_i + \gamma_4 \mbox{\tt BMI}_i + \alpha_1 m_i(t) \bigl \},\\
M_2: && h_i(t) = h_0(t) \exp \bigl \{\gamma_1 \mbox{\tt RR}_i + \gamma_2 \mbox{\tt Age}_i + \gamma_3 \mbox{\tt Female}_i + \gamma_4 \mbox{\tt BMI}_i + \alpha_1 m_i(t) + \alpha_2 m_i'(t)  \bigl \},\\
M_3: && h_i(t) = h_0(t) \exp \Bigl \{\gamma_1 \mbox{\tt RR}_i + \gamma_2 \mbox{\tt Age}_i + \gamma_3 \mbox{\tt Female}_i + \gamma_4 \mbox{\tt BMI}_i + \alpha_1 \int_0^t m_i(s) ds  \Bigl \},\\
M_4: && h_i(t) = h_0(t) \exp \bigl (\gamma_1 \mbox{\tt RR}_i + \gamma_2 \mbox{\tt Age}_i + \gamma_3 \mbox{\tt Female}_i + \gamma_4 \mbox{\tt BMI}_i + \alpha_1 b_{i0} + \alpha_2 b_{i1} + \alpha_3 b_{i2} + \alpha_4 b_{i3} \bigl ),
\end{eqnarray*}
where the baseline hazard is approximated with B-splines, i.e.,
\[
\log h_0(t) = \gamma_{h_0,0} + \sum \limits_{q = 1}^Q \gamma_{h_0,q} B_q(t, \bv),
\]
with five internal knots placed at the corresponding percentiles of the observed event times, and {\tt Female} denotes the dummy variable for females. The estimation of these models was based on a Bayesian approach and an MCMC algorithm with a single chain of 115,000 iterations from which we discarded the first 15,000 samples as burn-in. For all parameters we took standard prior distributions \citep{ibrahim.et.al:01, lesaffre.lawson:12}. In particular, for the vector of fixed effects of the longitudinal submodel $\bfbeta$, the regression parameters of the survival model $\bfgamma$, the vector of spline coefficients for the baseline hazard $\bfgamma_{h_0}$, and for the association parameter $\alpha$ we used independent univariate diffuse normal priors. For the variance of the error terms $\sigma^2$ we take an inverse-Gamma prior, while for covariance matrices we assumed an inverse Wishart prior. All computations have been performed in \R\, (version 3.0.1) using package \textbf{JMbayes} \citep[version 0.4-1;][]{JMbayes} and \textsf{WinBUGS} (version 1.4.3). Trace plots did not show any alarming indications of convergence failure while auto-correlation plots showed relatively good mixing of the chains. Tables~\ref{Tab:Res-Y} and \ref{Tab:Res-T} show estimates and the corresponding 95\% credible intervals for the parameters in the longitudinal and survival submodels, respectively. We observe that the parameter estimates in the relative risk models show greater variability between the posited association structures (in particular between the time-dependent ($M_1$, $M_2$, and $M_3$) and the time-independent parameterizations ($M_4$)) than the parameters in the linear mixed models. However, we should note that the interpretation of the regression coefficients $\bfgamma$ is not the same in the four survival submodels because we condition on different components of the longitudinal process.
\begin{table}[ht]
\scriptsize
\centering
\caption{Estimated coefficients and 95\% credible intervals for the parameters of the longitudinal submodels based on the four joint models fitted to the Aortic Valve dataset.}
\begin{tabular}{lrrrrrrrr}
  \hline
  & \multicolumn{2}{c}{Value ($M_1$)} & \multicolumn{2}{c}{Value+Slope ($M_2$)} & \multicolumn{2}{c}{Area ($M_3$)} & \multicolumn{2}{c}{Shared RE ($M_4$)}\\
 & Est. & 95\% CI & Est. & 95\% CI & Est. & 95\% CI & Est. & 95\% CI \\
  \hline
  {\tt SI} & 3.41 & (3.055; 3.772) & 3.38 & (3.030; 3.718) & 3.41 & (3.047; 3.747) & 3.38 & (3.020; 3.733) \\
  {\tt RR} & 2.86 & (2.687; 3.028) & 2.84 & (2.671; 3.005) & 2.87 & (2.707; 3.038) & 2.85 & (2.682; 3.017) \\
  {\tt SI}:B-spln1 & 1.42 & (0.885; 1.997) & 1.51 & (0.978; 2.048) & 1.37 & (0.840; 1.898) & 1.59 & (1.065; 2.163) \\
  {\tt RR}:B-spln1 & 1.38 & (0.933; 1.815) & 1.38 & (0.927; 1.839) & 1.42 & (0.970; 1.859) & 1.54 & (1.096; 2.005) \\
  {\tt SI}:B-spln2 & 2.94 & (1.867; 4.087) & 3.19 & (2.149; 4.113) & 2.79 & (1.874; 3.836) & 3.62 & (2.563; 4.745) \\
  {\tt RR}:B-spln2 & 2.57 & (1.723; 3.510) & 2.87 & (1.978; 3.796) & 2.32 & (1.458; 3.158) & 2.97 & (2.089; 3.911) \\
  {\tt SI}:B-spln3 & 3.56 & (2.585; 4.742) & 3.81 & (2.804; 4.787) & 3.31 & (2.417; 4.353) & 4.44 & (3.389; 5.672) \\
  {\tt RR}:B-spln3 & 2.36 & (1.024; 3.736) & 2.81 & (1.433; 4.269) & 1.94 & (0.654; 3.211) & 2.92 & (1.574; 4.319) \\
  $\sigma$ & 0.57 & (0.542; 0.608) & 0.58 & (0.545; 0.613) & 0.58 & (0.545; 0.611) & 0.58 & (0.549; 0.617) \\
   \hline
\end{tabular}
\label{Tab:Res-Y}
\end{table}
\begin{table}[ht]
\centering
\scriptsize
\caption{Estimated coefficients and 95\% credible intervals for the parameters of the survival submodels (parameters $\bfgamma_{h_0}$ of the baseline hazard have been omitted) based on the four joint models fitted to the Aortic Valve dataset.}
\begin{tabular}{lrrrrrrrr}
  \hline
  & \multicolumn{2}{c}{Value ($M_1$)} & \multicolumn{2}{c}{Value+Slope ($M_2$)} & \multicolumn{2}{c}{Area ($M_3$)} & \multicolumn{2}{c}{Shared RE ($M_4$)}\\
 & Est. & 95\% CI & Est. & 95\% CI & Est. & 95\% CI & Est. & 95\% CI \\
  \hline
  {\tt RR} & 0.34 & ($-$0.040; 0.739) & 0.36 & ($-$0.056; 0.790) & 0.29 & ($-$0.104; 0.688) & $-$0.01 & ($-$1.217; 1.112) \\
  {\tt Age} & 0.01 & ($-$0.001; 0.028) & 0.02 & (0.002; 0.034) & 0.00 & ($-$0.010; 0.018) & 0.06 & (0.022; 0.106) \\
  {\tt Female} & $-$0.15 & ($-$0.548; 0.225) & $-$0.12 & ($-$0.543; 0.275) & $-$0.13 & ($-$0.509; 0.243) & $-$0.45 & ($-$1.624; 0.567) \\
  {\tt BMI} & $-$0.07 & ($-$0.130; $-$0.019) & $-$0.08 & ($-$0.140; $-$0.023) & $-$0.06 & ($-$0.111; $-$0.002) & $-$0.21 & ($-$0.349; $-$0.095) \\
  $\alpha_1$ & 0.37 & (0.235; 0.496) & 0.28 & (0.106; 0.433) & 0.03 & (0.006; 0.044) & $-$0.47 & ($-$2.414; 1.543) \\
  $\alpha_2$ &  &  & 1.47 & ($-$0.261; 3.205) &  &  & $-$4.14 & ($-$7.931; $-$1.302) \\
  $\alpha_3$ &  &  &  &  &  &  & 0.97 & ($-$0.630; 3.130) \\
  $\alpha_4$ &  &  &  &  &  &  & 2.53 & (0.972; 4.383) \\
  \hline
  \end{tabular}
\label{Tab:Res-T}
\end{table}

To assess the predictive ability of the four joint models and compare them with the landmark approach we consider the time interval $[t = 7.5, u = 9.5]$ years. The reason for choosing this interval is twofold. First, by time $t = 7.5$ years 75\% of aortic gradient measurements have been recorded, and hence we have sufficient longitudinal information, and second, a two-year interval is considered a medically relevant time frame within which we would like to obtain accurate predictions of prognosis. For the 207 patients still at risk at $7.5$ years we fitted three Cox models with corresponding association structures to the joint models defined above (except from the random effects association structure), i.e.,
\begin{eqnarray*}
M_5: && h_i(u-7.5) = h_0(u-7.5) \exp \bigl \{\gamma_1 \mbox{\tt RR}_i + \gamma_2 \mbox{\tt Age}_i + \gamma_3 \mbox{\tt Female}_i + \gamma_4 \mbox{\tt BMI}_i + \alpha_1 \tilde y_i(7.5) \bigr \},\\
M_6: && h_i(u-7.5) = h_0(u-7.5) \exp \bigl \{\gamma_1 \mbox{\tt RR}_i + \gamma_2 \mbox{\tt Age}_i + \gamma_3 \mbox{\tt Female}_i + \gamma_4 \mbox{\tt BMI}_i\\
&& \hspace*{6cm} + \; \alpha_1 \tilde y_i(7.5) + \alpha_2 \tilde y_i'(7.5) \bigr \},\\
M_7: && h_i(u-7.5) = h_0(u-7.5) \exp \Bigl \{\gamma_1 \mbox{\tt RR}_i + \gamma_2 \mbox{\tt Age}_i + \gamma_3 \mbox{\tt Female}_i + \gamma_4 \mbox{\tt BMI}_i\\
&& \hspace*{6cm} + \; \alpha_1 \sum \limits_{s = 0}^{7.5} y_i(s) \Delta s \Bigr \},
\end{eqnarray*}
where $u > 7.5$, variable $\tilde y_i(7.5)$ denotes the last available square root aortic gradient value of each patient before year $7.5$,  $\tilde y_i'(7.5)$ denotes the slope defined from the last two available measurements, and $\sum_{0 \leq s \leq 7.5} y_i(s) \Delta s$ denotes the area under the step function defined from the observed square root aortic gradient measurements up to $7.5$ years. The parameter estimates and confidence intervals of these Cox models are presented in Table~\ref{Tab:Res-Cox}.
\begin{table}[ht]
\centering
\caption{Estimated coefficients and 95\% confidence intervals for the parameters in the Cox models fitted to the patients at risk at $t = 7.5$ years.}
\begin{tabular}{lrrrrrr}
  \hline
  & \multicolumn{2}{c}{Value ($M_5$)} & \multicolumn{2}{c}{Value+Slope ($M_6$)} & \multicolumn{2}{c}{Area ($M_7$)}\\
 & Est. & 95\% CI & Est. & 95\% CI & Est. & 95\% CI \\
  \hline
{\tt RR} & 0.42 & ($-$0.087; 0.930) & 0.42 & ($-$0.085; 0.927) & 0.39 & ($-$0.136; 0.907) \\
  {\tt Age} & $-$0.01 & ($-$0.025; 0.012) & $-$0.01 & ($-$0.024; 0.014) & $-$0.01 & ($-$0.026; 0.011) \\
  {\tt Female} & $-$0.17 & ($-$0.678; 0.347) & $-$0.16 & ($-$0.672; 0.352) & $-$0.15 & ($-$0.669; 0.363) \\
  {\tt BMI} & 0.02 & ($-$0.046; 0.093) & 0.03 & ($-$0.042; 0.097) & 0.03 & ($-$0.044; 0.094) \\
  $\alpha_1$ & 0.02 & ($-$0.187; 0.224) & $-$0.01 & ($-$0.221; 0.199) & $-$0.01 & ($-$0.047; 0.031) \\
  $\alpha_2$ &  &  & 0.25 & ($-$0.164; 0.669) &  &  \\
  \hline
\end{tabular}
\label{Tab:Res-Cox}
\end{table}
We evaluate both discrimination and calibration using the predictive accuracy measures presented in Section~\ref{Sec:PredAcc}, namely $\widehat{\mbox{PE}}(9.5|7.5)$, $\mbox{\ipeHat}(9.5|7.5)$, $\mbox{\aucHat}(9.5|7.5)$ and $\mbox{C}_{dyn}^{\Delta t=2}$. For the first two the absolute loss function was used, and the calculation of $\widehat{\mbox{C}}_{dyn}^{\Delta t=2}$ was based on the interval $[0, 15]$ years, with upper limit marking the 60\% percentile of the event times distribution. The estimates of these measures are presented in Table~\ref{Tab:AccMeas}.
\begin{table}[ht]
\centering
\begin{tabular}{lcccc}
  \hline
 & $\widehat{\mbox{PE}}(9.5|7.5)$ & $\mbox{\ipeHat}(9.5|7.5)$ & $\mbox{\aucHat}(9.5|7.5)$ & $\widehat{\mbox{C}}_{dyn}^{\Delta t=2}$ \\
  \hline
$M_1$: JM value & 0.1732 & 0.0904 & 0.6106 & 0.6433 \\
  $M_2$: JM value+slope & 0.1647 & 0.0855 & 0.5958 & 0.6592 \\
  $M_3$: JM area & 0.1525 & 0.0802 & 0.6090 & 0.5419 \\
  $M_4$: JM shared RE & 0.1133 & 0.0586 & 0.5755 & 0.6201 \\
   \hline
$M_5: \mbox{Cox}_{LM}$ value & 0.1888 & 0.1032 & 0.5587 & 0.6338 \\
  $M_6: \mbox{Cox}_{LM}$ value+slope & 0.1877 & 0.1025 & 0.5300 & 0.6238 \\
  $M_7: \mbox{Cox}_{LM}$ area & 0.1885 & 0.1031 & 0.5739 & 0.5930 \\
   \hline
\end{tabular}
\caption{Predictive performance measures for the Aortic Valve dataset under the four joint models and the landmark approach based on Cox models with the analogous functional forms. For $\widehat{\mbox{PE}}(9.5|7.5)$ and $\mbox{\ipeHat}(9.5|7.5)$ the absolute loss function was used. $\widehat{\mbox{C}}_{dyn}^{\Delta t=2}$ has been calculated in the interval $[0, 15]$ years.}
\label{Tab:AccMeas}
\end{table}
With respect to accuracy we observe that joint model $M_4$ with the shared random-effects parameterization has the smallest prediction error, followed by the other three joint models and the three Cox models using the landmark approach. This is in terms of both accuracy of prediction at year 9.5 and the weighted average of prediction errors in the interval $[7.5, 9.5]$. With respect to discriminative capability we observe that joint models $M_1$ and $M_2$ can best discriminate between patients followed by the landmark approach and the other two joint models. The overall winner could be deemed joint model $M_4$, which has the best accuracy and respectable discriminative capability compared to the models that offer the best discrimination. A comparison between the landmark approach and joint modeling in this particular dataset, and in particular when we compare the same parameterization (i.e., models $M_1$ vs. $M_5$, $M_2$ vs. $M_6$ and $M_3$ vs. $M_7$), reveals that the joint models perform better in terms of both accuracy and discrimination.


\section{Simulations} \label{Sec:Simul}
\subsection{Design} \label{Sec:Simul-Design}
We performed a series of simulations to landmarking with joint models in the context of dynamic predictions. The design of our simulation study is motivated by the set of joint models fitted to the Aortic Valve dataset in Section~\ref{Sec:Appl}. In particular, we assume 300 patients who have been followed-up for a period of 19 years, and were planned to provide longitudinal measurements at baseline and afterwards at nine random follow-up times. For the longitudinal process, and similarly to the model fitted in the Aortic Valve dataset, we used natural cubic splines of time with two internals knots placed at 2.1 and 5.5 years, and boundary knots placed at baseline and 19 years, i.e., the form of the model is as follows
\begin{eqnarray*}
y_i(t) & = & \beta_1 \mbox{\tt Trt0}_i + \beta_2 \mbox{\tt Trt1}_i + \beta_3 \{B_1(t, \bflambda) \times \mbox{\tt Trt0}_i\} + \beta_4 \{B_1(t, \bflambda) \times \mbox{\tt Trt1}_i\}\\
&& \hspace{0.5cm} + \, \beta_5 \{B_2(t, \bflambda) \times \mbox{\tt Trt0}_i\} + \beta_6 \{B_2(t, \bflambda) \times \mbox{\tt Trt1}_i\}\\
&& \hspace{0.5cm} + \, \beta_7 \{B_3(t, \bflambda) \times \mbox{\tt Trt0}_i\} + \beta_8 \{B_3(t, \bflambda) \times \mbox{\tt Trt1}_i\}\\
&& \hspace{0.5cm} + \, b_{i0} + b_{i1} B_1(t, \bflambda) + b_{i2} B_2(t, \bflambda) + b_{i3} B_3(t, \bflambda) + \eps_i(t),
\end{eqnarray*}
where $B_n(t, \bflambda)$ denotes the B-spline basis for a natural cubic spline with $\bflambda = (0, 2.1, 5.5, 19)$, {\tt Trt0} and  {\tt Trt1} are the dummy variables for the two treatment groups, $\eps_i(t) \sim \mathcal N (0, \sigma^2)$ and $\bb_i \sim \mathcal N (\mathbf 0, \bD)$ with $\bD$ taken to be diagonal.

For the survival process, we have assumed four scenarios, each one corresponding to a different functional form for the association structure between the two processes. Motivated by the arguments set forth in Section~\ref{Sec:DynPred-Comp}, we simulated survival data under the joint modeling framework (i.e., not assuming that the biomarker's levels are constant between the visit times). More specifically, \begin{eqnarray*}
\mbox{Scenario I:} && h_i(t) = h_0(t) \exp \bigl \{\gamma_0 + \gamma_1 \mbox{\tt Trt1}_i + \alpha_1 m_i(t) \bigl \},\\
\mbox{Scenario II:} && h_i(t) = h_0(t) \exp \bigl \{\gamma_0 + \gamma_1 \mbox{\tt Trt1}_i + \alpha_1 m_i(t) + \alpha_2 m_i'(t)  \bigl \},\\
\mbox{Scenario III:} && h_i(t) = h_0(t) \exp \bigl \{\gamma_0 + \gamma_1 \mbox{\tt Trt1}_i + \alpha_1 \int_0^t m_i(s) ds  \bigl \},\\
\mbox{Scenario IV:} && h_i(t) = h_0(t) \exp \bigl (\gamma_0 + \gamma_1 \mbox{\tt Trt1}_i + \alpha_1 b_{i0} + \alpha_2 b_{i1} + \alpha_3 b_{i2} + \alpha_4 b_{i3} \bigl ),
\end{eqnarray*}
with $h_0(t) = \sigma_t t^{\sigma_t - 1}$, i.e., the Weibull baseline hazard. The values for the regression coefficients in the longitudinal and survival submodels, the variance of the error terms of the mixed model, the covariance matrix for the random effects, and the scale of the Weibull baseline risk function are given in Appendix~\ref{App:SimSet}, and have been chosen such that the distribution of the event times and the distribution of the follow-up longitudinal measurements were comparable across scenarios. Censoring times were simulated from a uniform distribution in $(0, t_{C})$ with $t_{C}$ set to result in about 45\% censoring in each scenario. For each scenario we simulated 200 datasets.


\subsection{Results} \label{Sec:Simul-Results}
Mimicking the real-life use of a prognostic model, and to assess any potential overfitting issues, the comparison between the landmark and joint modeling approaches is based on subjects who were not used in fitting the corresponding models. More specifically, under each scenario and for each simulated dataset, we randomly excluded ten subjects whose event times were censored. For these subjects we set as landmark time the time point of their last longitudinal measurement, and produce survival probabilities from that point onwards to the end of the follow-up. Under the landmark approach these probabilities are based on the following relative risk models fitted to the remaining subjects:
\begin{eqnarray*}
LM_1: && h_i(u-t_{LM}) = h_0(u-t_{LM}) \exp \bigl \{\gamma_0 + \gamma_1 \mbox{\tt Trt1}_i + \alpha_1 \tilde y_i(t_{LM}) \bigr \},\\
LM_2: && h_i(u-t_{LM}) = h_0(u-t_{LM}) \exp \bigl \{\gamma_0 + \gamma_1 \mbox{\tt Trt1}_i + \alpha_1 \tilde y_i(t_{LM}) + \alpha_2 \tilde y_i'(t_{LM}) \bigr \},\\
LM_3: && h_i(u-t_{LM}) = h_0(u-7.5) \exp \Bigl \{\gamma_0 + \gamma_1 \mbox{\tt Trt1}_i + \alpha_1 \sum \limits_{s = 0}^{t_{LM}} y_i(s) \Delta s \Bigr \},
\end{eqnarray*}
where $t_{LM}$ denotes the landmark time, and as before, $\tilde y_i(t_{LM})$ denotes the last available measurement of subject $i$ before $t_{LM}$, $\tilde y_i'(t_{LM})$ denotes the slope defined from the last two available measurements, and $\sum \limits_{s = 0}^{t_{LM}} y_i(s) \Delta s$ the area under the step function defined from the observed longitudinal responses up to $t_{LM}$. Similarly, we also fitted four joint models to the remaining 290 subjects, with the same longitudinal submodel as the one we simulated from, and survival submodels:
\begin{eqnarray*}
JM_1: && h_i(t) = h_0(t) \exp \bigl \{\gamma_0 + \gamma_1 \mbox{\tt Trt1}_i + \alpha_1 m_i(t) \bigr \},\\
JM_2: && h_i(t) = h_0(t) \exp \bigl \{\gamma_0 + \gamma_1 \mbox{\tt Trt1}_i + \alpha_1 m_i(t) + \alpha_2 m_i'(t) \bigr \},\\
JM_3: && h_i(t) = h_0(t) \exp \Bigl \{\gamma_0 + \gamma_1 \mbox{\tt Trt1}_i + \alpha_1 \int_0^t m_i(s) \, ds \Bigr \},\\
JM_4: && h_i(t) = h_0(t) \exp \bigl(\gamma_0 + \gamma_1 \mbox{\tt Trt1}_i + \alpha_1 b_{i0} + \alpha_2 b_{i1} + \alpha_3 b_{i2} + \alpha_4 b_{i3} \bigr ),\\
\end{eqnarray*}
based on which survival probabilities were derived. Due to the fact that our aim here is to investigate the impact on predictions of the underlying differences between landmarking and joint modeling, as explained in Section~\ref{Sec:DynPred-Comp}, in both approaches the baseline hazard is assumed of the Weibull form, i.e., $h_0(t) = \sigma_t t^{\sigma_t - 1}$ with $\sigma_t$ denoting the shape parameter and the intercept term $\gamma_0$ the log scale parameter.

\begin{figure}[!h]
\includegraphics[width = \textwidth]{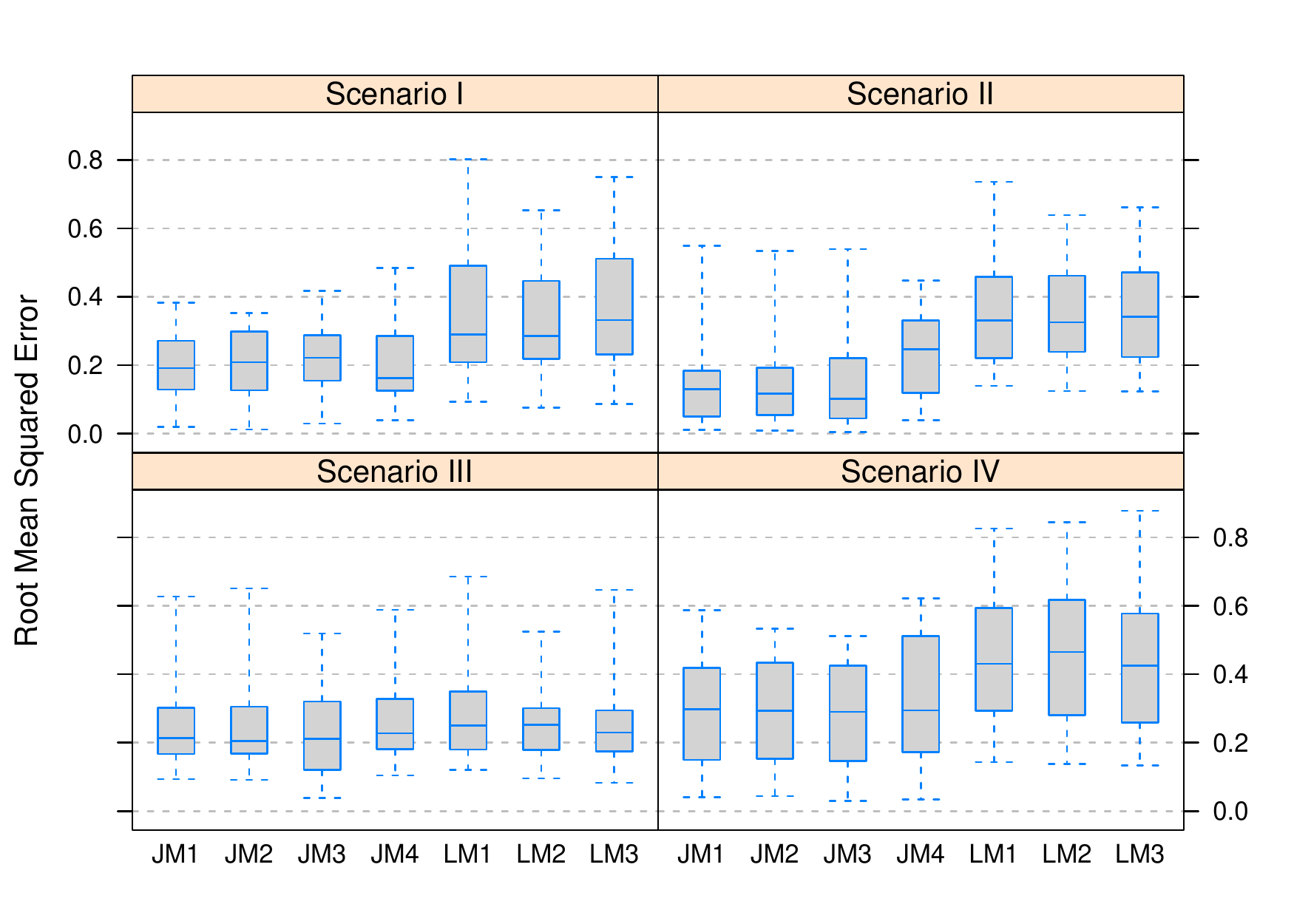}
\caption{Simulation results under the four scenarios based on 200 datasets. Each boxplot shows the distribution of the root mean squared predictions error of the corresponding model to compute predictions versus the gold standard.} \label{Fig:SimulResults}
\end{figure}

Based on the seven models, predictions were calculated for each of the ten subjects we have originally excluded, at ten equidistant time points between their last available longitudinal measurement and the end of follow-up. To evaluate the accuracy of these predicted survival probabilities we compared them with the gold standard survival probabilities, which are calculated as $S_j \bigl \{ u \mid \mathcal M_j(u, \bb_j), \bftheta \bigr \} / S_j \bigl \{ t \mid \mathcal M_j(t, \bb_j), \bftheta \bigr \}$, using the true parameter values and the true values of the random effects for the subjects we excluded. Hence, in each simulated dataset and for each of the ten subjects, we calculated root mean squared prediction errors (RMSEs) between the gold standard survival probabilities and the predictions under the seven models. The RMSEs over all the subjects from the 200 datasets are shown in Figure~\ref{Fig:SimulResults}.
The results suggest that the joint modeling approach seems to give more accurate predictions than landmarking. More noticeable are the differences in Scenarios I, II and IV while in Scenario III both approaches gave similarly accurate results.


\section{Discussion} \label{Sec:Disc}
In this work we have contrasted and compared two popular approaches, namely landmarking and joint modeling, for producing dynamically-updated predictions of survival probabilities with time-dependent covariates. Landmarking can effortlessly be implemented in practice but it makes strong assumptions regarding the path of the time-dependent covariates, which may be unrealistic for longitudinal biomarker measurements. On the other hand, joint modeling allows for greater flexibility in the attributes of time-dependent covariate process, but requires more modeling assumptions to achieve this and is generally more computationally intensive. Our simulation study and the analysis of the motivating Aortic Valve dataset have shown that, in general, there is a gain from considering the joint modeling approach instead of landmarking.

In our developments we have only focused on a single continuous longitudinal biomarker. However, often in practice and in order to obtain a more complete picture of the progression of a patient, several biomarkers are recorded, which could be of either continuous or categorical nature. In this more complex setting landmarking is advantageous because it is straightforward to include extra markers as baseline covariates in the linear predictor of the Cox model fitted to the patients at risk at the landmark point. On the contrary, the joint modeling approach requires a model specification for each marker. Mathematically and under the conditional independence assumptions \eqref{Eq:CondInd-I} and \eqref{Eq:CondInd-II} this relatively easily achieved by considering the framework of generalized linear mixed effects models \citep{breslow.clayton:93}. From the practical side, however, the dimensionality of the random effects may increase considerably, making joint models harder to fit. Previous and recent work by the first author is focused on resolving this problem by making use of Laplace approximations and efficient Gaussian quadrature rules \citep{rizopoulos.et.al:09, rizopoulos:12b}. In addition, in our analysis of the Aortic Valve dataset we have considered the composite event re-operation or death (whatever comes first), but for the treating physicians it could be of interest to have risk estimates separately for the two events. In this case we can extend both landmarking and joint modeling to the competing risks setting and derive estimates of the corresponding cumulative incidence functions. A general challenge when either or both of the two extensions (i.e., multiple longitudinal outcomes or multiple event times) are considered is the number of possible models. In particular, following the discussion in Section~\ref{Sec:FunForm} and the different possibilities we have in building the functional relationship between the longitudinal and time-to-event outcomes, it is evident that when we move to the multivariate setting, the choice of the appropriate parameterization for each longitudinal outcome and eventually for each competing risk becomes a demanding model-selection exercise.

Regarding the software implementation of the methodology presented in the paper, the landmark approach is readily available in all statistical software that fit Cox models. The fitting of joint models, the derivation of dynamic predictions (for the survival and longitudinal outcomes) and the calculation of the calibration and discrimination measures presented in Section~\ref{Sec:PredAcc} are implemented in the freely available \textsf{R} packages \textbf{JM} \citep{rizopoulos:10, rizopoulos:12} and \textbf{JMbayes} \citep{JMbayes}, which can be downloaded from CRAN at \url{http://cran.r-project.org/package=JM} and \url{http://cran.r-project.org/package=JMbayes}, respectively.


\appendix
\section{Simulation Settings} \label{App:SimSet}
For all simulation scenarios the parameter values that were used for the
longitudinal submodel were
\begin{itemize}
\item[] Fixed effects: $\beta_1 = 0.93$, $\beta_2 = -0.6$, $\beta_3 =
0.63$, $\beta_4 = 0.42$, $\beta_5 = 1.1$, $\beta_6 = 0.54$, $\beta_7 =
0.54$, and $\beta_8 = 0.55$;
\item[] Random effects diagonal covariance matrix: $\bD_{11} = 0.49$,
$\bD_{22} = 4.52$, $\bD_{33} = 2.33$, and $\bD_{44} = 1.52$;
\item[] Measurement error standard deviation: $\sigma = 2$.
\end{itemize}

For the survival submodels the parameters that were used to simulate
from each scenario are given in Table~\ref{Tab:Simulation}.

\begin{table}[ht]
\begin{center}
\begin{tabular}{crrrr}
\hline
& \multicolumn{4}{c}{Scenario}\\
& I & II & III & IV\\
\hline
$\gamma_0$ & $-$6.73 & $-$6.73 & $-$6.73 & $-$6.73 \\
$\gamma_1$ &    0.41 &    0.41 &     0.41 &    0.41 \\
$\alpha_1$ &    0.7 &    0.05 &     0.08 & $-$0.3 \\
$\alpha_2$ &         & 3.3 &          & $-$0.8 \\
$\alpha_3$ &         &         &          &    0.3 \\
$\alpha_4$ &         &         &          &    0.8 \\
$\sigma_t$ &    1.65 &    1.65 &     1.65 &    1.60 \\
\hline
\end{tabular}
\caption{Parameter values for the survival submodels under the four
simulation scenarios.}
\label{Tab:Simulation}
\end{center}
\end{table}


\section*{Acknowledgements}
\noindent We would like to thank the Interdisciplinary Centre for Mathematical and Computational Modeling from University of Warsaw for the possibility of using the Halo2 ICM supercomputer system.\\

\noindent {\bf{Conflict of Interest}}
\noindent {\it The authors have declared no conflict of interest.}


\bibliographystyle{biom}
\bibliography{CompParam}

\end{document}